\documentclass[12pt]{article}
\usepackage{amsmath,latexsym,epsfig,amssymb}

\def\bea{\begin{eqnarray}}
\def\eea{\end{eqnarray}}

\newcommand\be{\begin{equation}}
\newcommand\ee{\end{equation}}

\def\ma[#1,#2,#3,#4]  {{\left( \matrix{ #1  & #2 \cr
                                        #3  & #4 \cr } \right)}}

\begin{document}
\title{
{\vspace{-2cm} \normalsize
\hfill \parbox{40mm}{CERN/TH-99-376}\\
\hfill \parbox{40mm}{CPT-99/PE.3916}\\
\hfill \parbox{40mm}{LAPTH-Conf-771/99}}\\[25mm]
A numerical treatment of Neuberger's lattice Dirac operator\footnote{Talk given by K.J. at 
the ``Interdisciplinary Workshop on Numerical Challenges to Lattice QCD'', Wuppertal, August 22-24,1999}}
\author{
Pilar Hern\'andez\footnote{On leave from Departamento de F\'{\i}sica Te\'orica, Universidad de Valencia.},
Karl Jansen\footnote{Heisenberg Foundation Fellow}\\
               \\
CERN, 1211 Geneva 23, Switzerland\\
               \\
$\;$ and Laurent Lellouch\footnote{On leave from Centre de Physique 
Th\'eorique, CNRS Luminy, F-13288 Marseille Cedex 9, France.}\\
                                    \\
LAPTH, Chemin de Bellevue, B.P. 110, \\ F74941 Annecy-le-Vieux Cedex, France
}
%
\maketitle

\vspace{-1cm}
\begin{abstract}
We describe in some detail our numerical treatment of Neuberger's lattice Dirac operator
as implemented in a practical application. We discuss the improvements we have found to
accelerate the numerical computations and give an estimate of the expense
when using
this operator in practice.
\end{abstract}

\pagebreak

\section{Lattice formulation of QCD}

Today, we believe that the world of quarks and gluons is described theoretically by
quantum chromodynamics (QCD). This model shows a number
of non-perturbative aspects that cannot be adequately addressed 
by approximation schemes such as perturbation theory. 
The only way to evaluate QCD, addressing both
its perturbative and non-perturbative aspects {\em at the same time}, 
is lattice QCD. In this approach the theory is put on a 4-dimensional
Euclidean space-time lattice of finite physical length $L$,  
with a non-vanishing value of the 
lattice spacing $a$. Having only a finite number of grid points,
physical quantities can be computed numerically
by solving a high-dimensional integral by Monte Carlo methods,
making use of importance sampling.

The introduction of a lattice spacing regularizes    
the theory and is an intermediate step in the computation of
physical observables. 
Eventually, the regularization has to be removed and 
the value of the lattice spacing has to be sent
to zero to reach the target theory, i.e. continuum QCD. In fact, in conventional
formulations of lattice QCD \cite{wilson}, the introduction of the lattice spacing
renders the theory on the lattice somewhat different from the 
continuum analogue and a number of properties of the continuum theory
are only very difficult and cumbersome to establish in the lattice
regularized theory. One of the main reasons for this difficulty is that
in conventional lattice QCD the regularization breaks a particular symmetry of the
continuum theory,  
which plays a most important role there, namely chiral symmetry. 

However, the last few years have seen a major breakthrough in that we now have 
formulations of lattice QCD that have an exact lattice chiral symmetry \cite{luscher}. 
In this approach, 
many properties of continuum QCD are preserved 
{\em even at a non-vanishing value of the lattice spacing} \cite{hasen,hasen3,neu,luscher,chandra}. 
This development followed the
rediscovery \cite{hasen2} of the so-called Ginsparg--Wilson (GW) relation \cite{gw} which
is fulfilled by
any operator with the exact lattice chiral symmetry of \cite{luscher}. 
It is not the aim of this contribution to discuss the physics consequences
of the GW relation. We have to refer the interested reader to reviews \cite{nieder,blum} about these topics.
Here we would like to
discuss the {\em numerical} treatment of a particular lattice operator that 
satisfies the GW relation, namely Neuberger's solution \cite{neu}. 
This solution has a complicated structure and is challenging to implement 
numerically. 
Thus, the large theoretical advantage of an operator satisfying the
GW relation must be weighed against the very demanding computational effort
required to implement it.

This contribution is organized as follows. 
After discussing Neuberger's lattice Dirac operator we want to show how we
evaluated the operator in our practical application \cite{ourwork} and what kind of 
improvements we found to accelerate the numerical computations. 
For alternative ideas for improvements, see the contributions of 
H. Neuberger \cite{neuwork} and A. Borici \cite{artan} to this workshop. 
We finally give some estimates
of the computational expense of using Neuberger's operator. 

\section{Neuberger's lattice Dirac operator}

The operator we have used acts on fields
(complex vectors) 
$\Phi(x)$ where $x=(x_0,x_1,x_2,x_3)$ and the $x_\mu, \mu=0,1,2,3$, 
are integer numbers denoting a 4-dimensional grid point in a lattice
of size $N^4$ with $N=L/a$. The fields $\Phi(x)$ carry in addition
a ``colour'' index $\alpha=1,2,3$ as well as a ``Dirac'' index 
$i=1,2,3,4$. Hence, $\Phi$ is a $N^4\cdot 3\cdot 4$ 
complex vector. 

In order to reach 
the expression for Neuberger's operator we first introduce the
matrix $A$
\begin{equation}
\label{defofA}
A = 1 + s - \frac{a}{2}\left\{\gamma_\mu\left(\nabla_\mu^{*}+\nabla_\mu\right) 
      - a \nabla_\mu^{*}\nabla_\mu \right\}\; ,
\end{equation}
where $\nabla_\mu$ and $\nabla_\mu^*$ are the nearest-neighbour forward
and backward derivatives, the precise definition of which can be found 
in the Appendix. The parameter $s$ is to be taken in the range of $|s|<1$
and serves to optimize the localization properties \cite{hjl} of Neuberger's operator,
which is then given by
\begin{equation}
\label{defofneu}
D = \frac{1}{a}\left\{1-A\left(A^\dagger A\right)^{-1/2}\right\}.
\end{equation}
Through the appearance of the square root in eq.~(\ref{defofneu}), 
all points on the lattice are connected with one another, giving rise
to a very complicated, multi-neighbour action. However, the application of 
$D$ to a vector $\Phi$ will only contain applications of $A$ or $A^\dagger A$
on this vector. Since these matrices are sparse, as only nearest-neighbour
interactions are involved, we will never have to store the whole matrix. 

In the computation of 
physical quantities, the inverse of $D$, applied to a given 
vector, is generically needed. Hence one faces the problem of having to compute 
a vector $X=D^{-1}\eta$, with $\eta$ a prescribed vector (the ``source'')
as required by the particular problem under investigation. 
Fortunately, 
a number of efficiently working algorithms 
for computing $X=D^{-1}\eta$ are known, such as conjugate gradient, BiCGstab, 
or variants thereof \cite{book}.  
In conventional approaches to lattice QCD an operator $\tilde{D}$ is used that is very similar 
to the matrix $A$ in eq.~(\ref{defofA}). 
Computing the vector $\tilde{X}=\tilde{D}^{-1}\eta$ requires a number 
$n_{\rm iter}$ of iterations of some particular method, say BiCGstab. 
Employing Neuberger's operator $D$ in computing 
$X=D^{-1}\eta$, it turns out that the number of iterations needed is
of the same order of magnitude as when using $\tilde{D}$. 
At the same time, in each of these iterations, the square root has to be
evaluated. 
When this is done by some polynomial approximation, 
it is found that the required degree of this polynomial 
is roughly of the same order as the number of iterations needed for computing
the vector $X$. 
Hence, with respect to the conventional case, 
the numerical effort is squared and 
the price to pay for using  
the operator $D$ is high. 

On the other hand, any solution of the Ginsparg--Wilson relation gives us a tool
by which particular problems in lattice QCD can be studied, which would be extremely
hard to address with conventional approaches. It is for these cases that
the large numerical effort is justified, but clearly, we would like to have 
clever ideas coming from areas such as Applied Mathematics, to decrease the numerical
expense or even overcome this bottleneck.

\section{Approximation of $\left(A^\dagger A\right)^{-1/2}$}

\begin{figure}
\vspace{0.0cm}
\begin{center}
\psfig{file=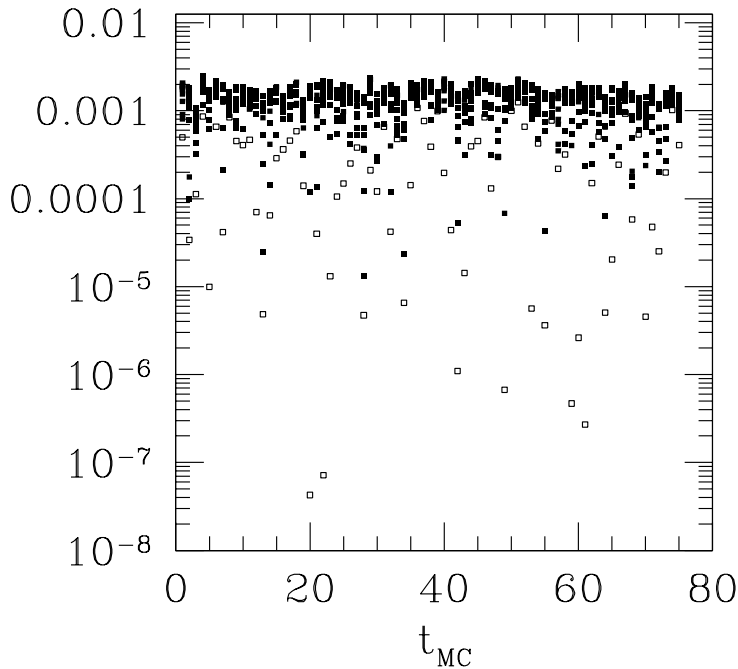, %
width=13cm,height=13cm}
\end{center}
\caption{ \label{fig:eigen_aa} 
Monte Carlo time evolution of the eleven lowest eigenvalues of $A^\dagger A$
at $\beta=5.85$. 
The lowest eigenvalue for each
configuration is the open square.
}
\end{figure}

For computing the square root that appears in eq.~(\ref{defofneu}), we have chosen 
a Chebyshev approximation \cite{numrec} by constructing a polynomial $P_{n,\epsilon}(x)$
of degree $n$, which has an exponential convergence rate in the 
interval $x\in [\epsilon,1]$. Outside this interval, convergence
is still found but it will not be exponential.
The advantages of using this Chebyshev approximation are the well-controlled
exponential fit accuracy as well as the possibility of having 
numerically very stable recursion relations \cite{foxy} to construct the
polynomial, allowing for large degrees. 
In order to have an optimal approximation,
it is desirable to know the lowest and the highest eigenvalue 
of $A^\dagger A$. 
A typical example of the eigenvalues of $A^\dagger A$ is shown
in fig.~\ref{fig:eigen_aa}, where we show the 11 lowest eigenvalues as 
obtained on a number of configurations using the Ritz functional method \cite{cg}. 
There is a wide spread and very low-lying eigenvalues appear. 
Choosing $\epsilon$ to be the value of the lowest of these eigenvalues would 
result in a huge degree $n$ of the polynomial $P_{n,\epsilon}$. We therefore
computed ${\rm O}(10)$ lowest-lying eigenvalues of
$A^\dagger A$ as well as their eigenfunctions and projected them
out of the matrix $A^\dagger A$. 
The approximation is 
then only performed for the matrix with a reduced condition number, 
resulting in a substantial decrease of the degree of the polynomial.
In addition, we computed the highest eigenvalue of $A^\dagger A$ 
and normalized the matrix $A$ such that $\| A^\dagger A\| \lesssim 1$.  

Since our work \cite{ourwork}, aiming at the physical question of spontaneous
chiral symmetry breaking in lattice QCD, has been one of the first of its
kind, we wanted to exclude possible systematic errors and demanded
a very high precision for the approximation to the square root:
\begin{equation}
\label{precofsq}
\| X - P_{n,\epsilon}(A^\dagger A) A^\dagger A P_{n,\epsilon}(A^\dagger A) X\|^2 /\|2X\|^2
       < 10^{-16}
\end{equation}
where $X$ is a gaussian random vector. In our practical applications we fixed this
precision beforehand and set $\epsilon$ to be the $11$th lowest eigenvalue of $A^\dagger A$.
This then determines the degree of the polynomial $n$ and hence
our approximation $D_n$ to the exact Neuberger operator $D$. 
We checked that the precision we required for the approximation of the
square root is directly related to the precision by which the GW relation
itself is fulfilled. Choosing $n$ such that 
the accuracy 
in eq.~(\ref{precofgw}) is reached results in 
\begin{equation}
\label{precofgw}
\| \left[\gamma_5 D_n +D_n\gamma_5 - D_n\gamma_5D_n\right]X\|^2 /\|X\|^2 \approx 
        10^{-16}\; .
\end{equation}
In addition, we find that the deviations from the exact GW relation decrease exponentially fast
with increasing $n$. 

\section{The inverse of Neuberger's operator}

\begin{figure}
\vspace{0.0cm}
\begin{center}
\psfig{file=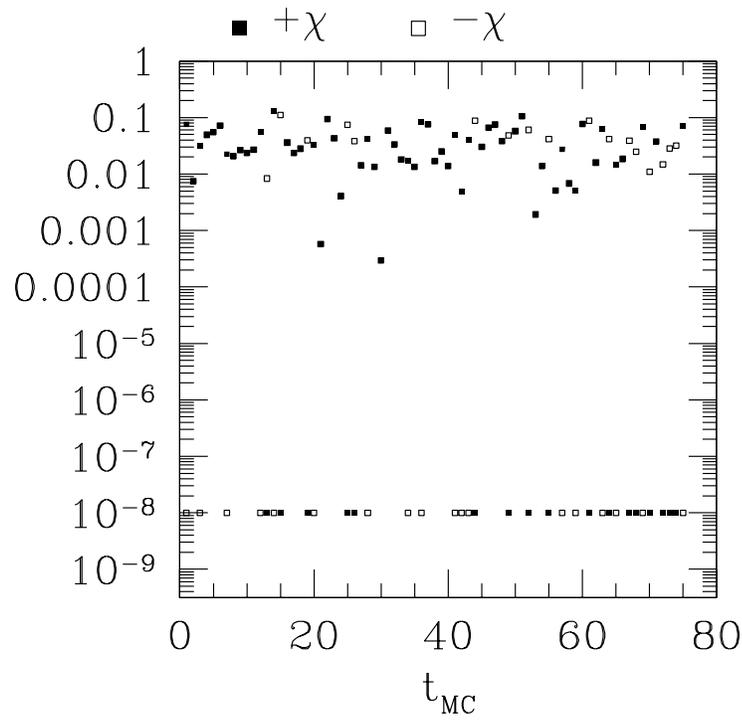, %
width=13cm,height=13cm}
\end{center}
\caption{ \label{fig:eigen_dd}
Monte Carlo time evolution of the lowest eigenvalue of $D_n^\dagger D_n$. 
The eigenvalues belong to given chiral sectors of the theory denoted
as $\pm\chi$ for chirality plus (full squares) and minus (open squares).
Data are obtained at $\beta=5.85$ choosing $s=0.6$.
Whenever there is a zero mode of $D_n^\dagger D_n$, the value of the 
lowest eigenvalue is set to $10^{-8}$.
}
\end{figure}

As mentioned above, in physics applications a vector
$D^{-1}\eta$ has to be computed, with $\eta$ a prescribed source 
vector. 
Not only is the computation of this vector very costly, there also appears 
to be a conceptual problem: in inspecting the lowest eigenvalue of
$D_n^\dagger D_n$, very small eigenvalues are often found 
as shown in 
fig.~\ref{fig:eigen_dd}. These very small eigenvalues belong to a given
chiral sector of the theory, i.e. their corresponding eigenfunctions $\chi$
are eigenfunctions of $\gamma_5$ with $\gamma_5\chi = \pm\chi$. 
In fact, these modes play an important physical role as they are associated
with topological sectors of the theory \cite{hasen3,hasen,luscher,neu}. 

As far as the practical applications are concerned, it is clear that in the presence
of such a small eigenvalue, the inversion of $D_n$ will be very costly, as the condition number
of the problem is then very high. 
In order to address this problem, we followed two strategies:

\begin{itemize} 
\item [ $(i)$ ] We compute the lowest eigenvalue of $D_n^\dagger D_n$ and its eigenfunction
                (using again the Ritz functional method \cite{cg})
                and if it is
                a zero mode --in which case it is also a zero mode of $D_n$--
                we project this mode out of $D_n$ and invert only the 
                reduced matrix; this is then well conditioned, as the very small
                eigenvalues appear to be isolated. 
                In this strategy, the knowledge of the eigenfunction must
                be very precise and an accuracy of approximating the square root
                as indicated in eq.~(\ref{precofsq}) is mandatory.
\item [ $(ii)$ ] Again we determine the lowest eigenvalue of $D_n^\dagger D_n$
                 and
                 the chirality of the corresponding zero mode, if there is any. 
                 We then make use of the fact
                 that $D_n^\dagger D_n$ 
                 commutes with $\gamma_5$. This allows us to perform the inversion 
                 in the chiral sector {\em without zero modes}. 
                 In this strategy, the accuracy demanded in eq.~(\ref{precofsq}) 
                 could be relaxed and this strategy, which essentially follows
                 ref.~\cite{florida}, is in general much less expensive than 
                 following strategy $(i)$.
\end{itemize}

However, even adopting strategy $(ii)$, solving the system $D_nX=\eta$ is 
still costly. We therefore tried two ways of improving on this.
We first note that instead of solving
\begin{equation}
\label{formulation1}
\left[1-A/\sqrt{A^\dagger A}\right] X =\eta
\end{equation}
we can equally well solve
\begin{equation}
\label{formulation2}
\left[A^\dagger-\sqrt{A^\dagger A}\right] X =A^\dagger\eta\; .
\end{equation}
In practice, however, we found no real advantage in using 
the formulation of eq.~(\ref{formulation2}). 
We have further considered two acceleration schemes.

\vspace{0.5cm}
\noindent {\em Scheme (a)}
\vspace{0.5cm}

We choose two different polynomials (now approximating 
$\sqrt{A^\dagger A}$ and not the inverse) $P_{n,\epsilon}$ and $P_{m,\epsilon}$, $m<n$,
such that 
\begin{equation}
P_{n,\epsilon} = P_{m,\epsilon} + \Delta           
\end{equation}
with $\Delta$ a ``small'' correction. Then we have 
\begin{eqnarray}
\left[ A^\dagger - P_{n,\epsilon}\right]^{-1} & = & \left[ A^\dagger - P_{m,\epsilon}-\Delta \right]^{-1}
                 \nonumber \\
& \approx & \left[ 1 + \left(A^\dagger - P_{m,\epsilon}\right)^{-1}\Delta\right]
                       \left(A^\dagger - P_{m,\epsilon}\right)^{-1}\; .
\end{eqnarray}

\noindent This leads us to the following procedure of solving $D_nX=\eta$:
\begin{itemize}
\item [ (1) ] first solve

\begin{equation}
\left(A^\dagger - P_{m,\epsilon}\right) Y =\eta\; ;
\end{equation}

\item [ (2) ] then solve

\begin{equation}
\left(A^\dagger - P_{m,\epsilon}\right) X_0 =\eta+\Delta Y\; ;
\end{equation}

\item [ (3) ] use  $X_0$ as a starting vector to finally solve  

\begin{equation}
\left(A^\dagger - P_{n,\epsilon}\right) X =\eta\; .
\end{equation}

\end{itemize}

\begin{figure}
\vspace{0.0cm}
\begin{center}
\psfig{file=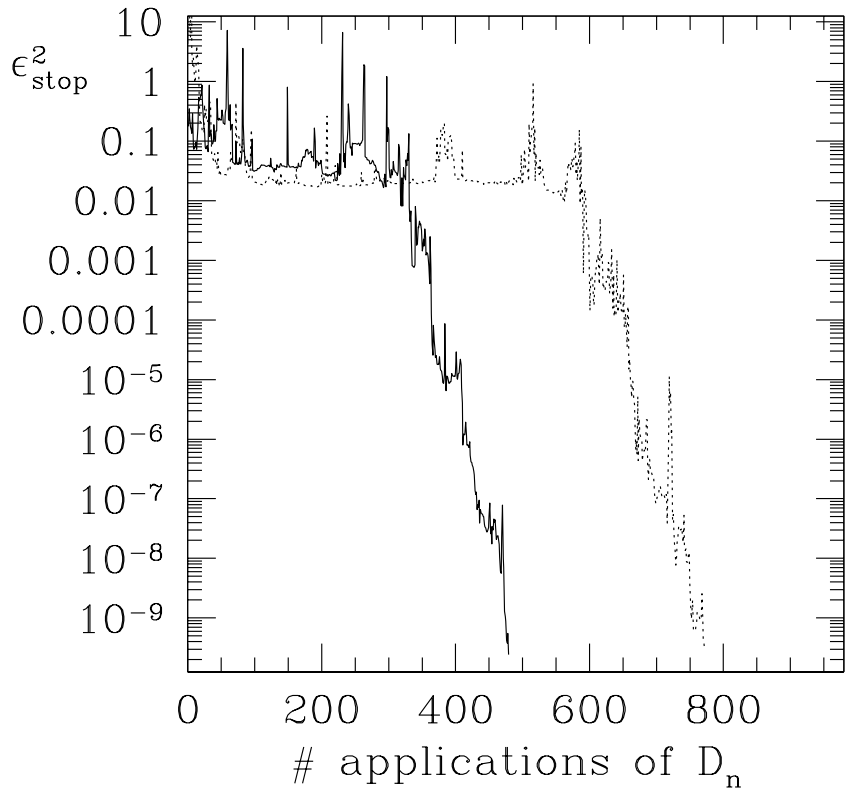, %
width=13cm,height=13cm}
\end{center}
\caption{ \label{fig:cg_nolg}
The residuum as a function of the number of applications of the
matrix $D_n$. 
The dotted line corresponds to a random starting vector. The solid 
line to a starting vector generated following scheme (a). 
}
\end{figure}

The generation of the starting vector $X_0$ in steps (1) and (2) 
is only a small overhead. In fig.~\ref{fig:cg_nolg}
we plot the 
relative residuum $\epsilon_{\rm stop}^2 = \|D_nX-\eta\|^2/\|X\|^2$ as a function of
the number of applications of $D_n$. In this case $n=100$ and 
$m=30$. We show the number of applications of the matrix $D_n$ for the
case of a random starting vector (dotted line) and the case where 
$X_0$ was generated according to the above procedure (solid line). 
The gain is of approximately a factor of two.

\vspace{0.5cm}
\noindent {\em Scheme (b)}
\vspace{0.5cm}

In the second approach, 
we use a sequence of polynomials to solve $D_nX=\eta$. 
To this end we first solve 
\begin{equation}
\left[A^\dagger-P_{m_1,\epsilon}\right] X_1 =A^\dagger \eta
\end{equation}
by choosing a polynomial $P_{m_1,\epsilon}$ and a stopping criterion
for the solver $\epsilon_{\rm stop}^{(1)}$ such that 
\begin{equation}
        m_1 < n, \;\epsilon_{\rm stop}^{(1)} > \epsilon_{\rm stop}\; .
\end{equation}
The value of $\epsilon_{\rm stop}^{(1)}$ is chosen such that it 
is roughly of the same order of magnitude as the error
that the polynomial of degree $m_1$ itself induces. 
The solution $X_1$ is then used as a starting vector for the next equation,       
employing a polynomial $P_{m_2,\epsilon}$ and stopping criterion 
$\epsilon_{\rm stop}^{(2)}$ with 
\begin{equation}
     m_1 < m_2 < n, \;\epsilon_{\rm stop}^{(1)} > \epsilon_{\rm stop}^{(2)} > \epsilon_{\rm stop}\; .
\end{equation}
This procedure is then repeated until we reach the desired polynomial
$P_{n,\epsilon}$ and stopping criterion $\epsilon_{\rm stop}$ to solve 
the real equation

\begin{equation}
\left[A^\dagger-P_{n,\epsilon}\right] X =A^\dagger \eta\; .
\end{equation}
           
As for scheme (a), we gain a factor of about two in the numerical effort. 
We finally remark that 
some first tests using the scheme proposed in \cite{artan} resulted in a similar performance gain as
the two schemes presented above.

In table~\ref{expense} we give a typical example of the expense 
of a simulation following strategy $(ii)$. 
We list both the cost of computing the lowest eigenvalue of $D_n^\dagger D_n$ in terms 
of the number of iterations to minimize the Ritz functional \cite{cg} and the
number of iterations to solve $D_nX=\eta$. 
In both applications, a polynomial
of degree $n$ is used to approximate the square root. 
The numbers in table~\ref{expense} indicate that a quenched calculation,
employing Neuberger's operator, leads to a computational cost that
is comparable with a dynamical simulation using conventional operators.

\begin{table}[htbp]
  \begin{center}
    \leavevmode
    \begin{tabular}[]{|c|c|c|c|}
\hline
$N$ & $n$ &  $n_{\rm ev}$ &  $n_{\rm invert}$ \\
\hline\hline
$8$  &  $190$ & $ 170$ &  $80$   \\
$10$ &  $250$ & $ 325$ &  $200$   \\
$12$ &  $325$ & $ 700$ &  $300$   \\
\hline
    \end{tabular}
\vspace{0.4cm}
\caption{
\label{expense}
$N$ is the number of lattice sites along a side of the hypercube;
$n$, the degree of polynomial;  $n_{\rm ev}$, the number
of iterations required to obtain the lowest eigenvalue of
$D_n^\dagger D_n$; and $n_{\rm invert}$, the number of iterations 
necessary to compute $X=D_n^{-1}\eta$.
        }
  \end{center}
\end{table}

\section{Conclusions}

The theoretical advance that an exact chiral symmetry brings
to lattice gauge theory is accompanied by the substantial increase in
numerical effort that is required to implement operators satisfying the
GW relation.  Thus, while the Nielsen--Ninomiya theorem has been
circumvented, the ``no free lunch theorem'' has not.  Whether
alternative formulations, such as domain wall fermions, can help in this
respect remains to be seen.

\section*{Appendix} 

We give here the explicit definitions needed in 
eq.~(\ref{defofA}). 
The forward and backward derivatives 
$\nabla_\mu,\;\nabla_\mu^*$ act on a vector $\Phi(x)$ as          
\begin{eqnarray*}
  \nabla_\mu \Phi(x)   &=& \frac{1}{a}\left[ U(x,\mu) \Phi(x+a\hat{\mu})- \Phi(x)\right] \nonumber \\
  \nabla_\mu^* \Phi(x) &=& \frac{1}{a}\left[ \Phi(x)-U(x-a\hat{\mu},\mu)^{-1} \Phi(x-a\hat{\mu})\right]\; ,
\end{eqnarray*}
where $\hat{\mu}$ denotes the unit vector in direction $\mu$. The (gauge) field $U(x,\mu) \in SU(3)$
lives on the links connecting lattice points $x$ and $x+a\hat{\mu}$ and acts on the
colour index $\alpha=1,2,3$ of the field $\Phi$.
Finally, the Dirac matrices $\gamma_\mu,\;\mu=0,1,2,3$ are hermitean $4\times 4$
matrices acting on the Dirac index $i$ of the field $\Phi$. Their explicit form is given by

\begin{equation}
\gamma_\mu = \left( \begin{array}{cc}
                  0 & e_\mu \\
                  e_\mu^\dagger & 0
                  \end{array} \right)
\end{equation}
with
\begin{equation}
e_0 = 1,\;\; e_k =-i\sigma_k
\end{equation}
and
\begin{equation}
\sigma_1 = \left( \begin{array}{cc}
                0 & 1 \\
                1 & 0
                \end{array} \right) \;\;\;\;
\sigma_2 = \left( \begin{array}{cc}
                0 & -i \\
                i & 0
                \end{array} \right) \;\;\;\;
\sigma_3 = \left( \begin{array}{cc}
                1 & 0 \\
                0 & -1
                \end{array} \right) \;.        
\end{equation}
With the choice of the $\gamma$ matrices given above, the matrix 
$\gamma_5=\gamma_0\gamma_1\gamma_2\gamma_3$ is diagonal and given by
\begin{equation}
\gamma_5 = \left( \begin{array}{cc}
                  1 & 0 \\
                  0 & -1
                  \end{array} \right)\; .
\end{equation}

We finally note that whenever repeated indices appear, they are summed over.


\begin{thebibliography}{99}

%
\bibitem{wilson} K.G. Wilson, Phys. Rev. {\bf D10} (1974) 2445.
%
\bibitem{hasen} P. Hasenfratz, V. Laliena and F. Niedermayer, Phys.Lett.{\bf B427} (1998) 125.
%
\bibitem{hasen3} P. Hasenfratz, Nucl. Phys. {\bf B525} (1998) 401.       
%
\bibitem{neu} H. Neuberger, Phys. Lett. {\bf B417} (1998) 141 and {\bf B427} (1998) 353.
%
\bibitem{luscher} M. L\"uscher, Phys. Lett. {\bf B428} (1998) 342.
%
\bibitem{chandra} S. Chandrasekharan, Phys. Rev. {\bf D60} (1999) 074503. 
%
\bibitem{hasen2} P. Hasenfratz, Nucl. Phys. B (Proc.Suppl.) {\bf 63A-C} (1998) 53.
%
\bibitem{gw} P.H. Ginsparg and K.G. Wilson, Phys. Rev. {\bf D25} (1982) 2649.
%
\bibitem{nieder} F. Niedermayer, Nucl. Phys. B (Proc.Suppl.) {\bf 73} (1999) 105. 
%
\bibitem{blum} T. Blum, Nucl.Phys.B (Proc.Suppl.) {\bf 73} (1999) 167.              
%
\bibitem{ourwork} 
P.~Hern\'andez, K.~Jansen and L.~Lellouch, CERN preprints CERN-TH/99-197, hep-lat/9907022 and
CERN-TH/99-273, hep-lat/9909026.
%
\bibitem{neuwork} H. Neuberger, hep-lat/9910040.
%
\bibitem{artan} A. Borici, hep-lat/9910045.
%
\bibitem{hjl} 
P. Hern\'andez, K. Jansen and M. L\"uscher, Nucl. Phys. {\bf B552} (1999) 363.
%
\bibitem{numrec}
W.~H.~Press, S.~A.~Teukolsky, W.~T.~Vetterling and B.~P.~Flannery,
{\em Numerical Recipes},
Second Edition, Cambridge University Press, Cambridge, 1992.
%
\bibitem{book} Y. Saad, {\em Iterative methods for sparse linear systems}, PWS Publishing Company,
Boston, 1996. 
%
\bibitem{foxy}
L.~Fox and I. B.~Parker, {\em Chebyshev polynomials in numerical analysis},
Oxford University Press, London, 1968.
%
\bibitem{florida} R.G. Edwards, U.M. Heller and R. Narayanan, Phys. Rev. {\bf D59} (1999) 094510.
%
\bibitem{cg}
B. Bunk, K. Jansen, M. L\"uscher and H. Simma, DESY report (September 1994);
T. Kalkreuter and H. Simma, Comp.Phys.Comm. {\bf 93} (1996) 33.
%
 

\end{thebibliography}
\end{document}